\documentclass[11pt,a4paper]{article}
\pdfoutput=1
\usepackage{amsmath,amssymb,amsfonts,dsfont}
\usepackage{url}
\usepackage{epsfig}
\usepackage{comment}
\usepackage{jheppub}

\newcommand{\ox}{\omega}
\newcommand{\ax}{\alpha}
\newcommand{\bx}{\beta}
\newcommand{\tox}{\tilde\omega}
\newcommand{\ex}{\epsilon}
\newcommand{\te}{\tilde \epsilon}
\newcommand{\tm}{\tilde \mu}
\newcommand{\cZ}{\mathcal{Z}}
\newcommand{\cG}{\mathcal{G}}
\newcommand{\barcZ}{\bar{\mathcal{Z}}}
\newcommand{\barcG}{\bar{\mathcal{G}}}
\newcommand{\cV}{\mathcal{V}}

\newcommand{\cK}{\mathcal{K}}
\newcommand{\tcK}{\tilde{\mathcal{K}}}

\newcommand{\cW}{{\cal W}}

\newcommand{\re}{\textrm{Re}}
\newcommand{\im}{\textrm{Im}}

\newcommand{\be}{\begin{equation}}
\newcommand{\ee}{\end{equation}}
\newcommand{\beq}{\begin{equation}}
\newcommand{\eeq}{\end{equation}}
\newcommand{\bea}{\begin{eqnarray}}
\newcommand{\eea}{\end{eqnarray}}
\newcommand{\ba}{\begin{array}}
\newcommand{\ea}{\end{array}}
\newcommand{\nn}{\nonumber}

\newcommand{\cN}{{\cal N}}

\newcommand{\barZ}{\bar{Z}}

\def\d{{\rm d}}
\def\p{\partial}

\title{$Spin(7)$ Compactifications and 1/4-BPS Vacua in Heterotic Supergravity}

\author[a]{Stephen Angus,}
\author[b,c]{Cyril Matti}
\author[d,e,f]{and Eirik E.~Svanes}

\affiliation[a]{Center for Theoretical Physics of the Universe, Institute for Basic Science (IBS),\\
	Daejeon 34051, Republic of Korea}
\affiliation[b]{Department of Mathematics, City University, London,\\
	Northampton Square, London EC1V 0HB, UK}
\affiliation[c]{Mandelstam Institute for Theoretical Physics, NITheP, and School of Physics,\\
University of the Witwatersrand, Johannesburg, WITS 2050, South Africa}
\affiliation[d]{Sorbonne Universit\'es, UPMC Univ Paris 06, UMR 7589, LPTHE, F-75005, Paris, France}
\affiliation[e]{CNRS, UMR 7589, LPTHE, F-75005, Paris, France}
\affiliation[f]{Sorbonne Universit\'es, Institut Lagrange de Paris, 98 bis Bd Arago, 75014 Paris, France\\}

\emailAdd{sangus88@ibs.re.kr}
\emailAdd{cyril.matti.1@city.ac.uk}
\emailAdd{esvanes@lpthe.jussieu.fr}

\abstract{We continue the investigation into non-maximally symmetric compactifications of the heterotic string. In particular, we consider compactifications where the internal space is allowed to depend on two or more external directions. For preservation of supersymmetry, this implies that the internal space must in general be that of a $Spin(7)$ manifold, which leads to a $1/4$-BPS four-dimensional supersymmetric perturbative vacuum breaking all but one supercharge. We find that these solutions allow for internal geometries previously excluded by the domain-wall-type solutions, and hence the resulting four-dimensional superpotential is more generic. In particular, we find an interesting resemblance to the superpotentials that appear in non-geometric flux compactifications of type II string theory. If the vacua are to be used for phenomenological applications, they must be lifted to maximal symmetry by some non-perturbative or higher-order effect.}

\preprint{
\begin{flushright}
CTPU-15-23
\end{flushright}
}

\begin{document}
\maketitle
\flushbottom


\section{Introduction}
Since the beginnings of string phenomenology in the mid 1980's \cite{Candelas:1985en}, the search for phenomenologically viable vacua from heterotic strings has been ongoing.  Most studies of string compactification have thus far focused on Calabi-Yau compactifications or slight generalizations thereof.  In the case of heterotic string theory it is comparatively easy to obtain Standard Model-like physics owing to the plethora of gauge bundles available in such a context, and a lot of progress has been made on this front in recent years, see e.g. \cite{Berglund:1995yu, Donagi:2004ub, Braun:2005bw, Braun:2005nv, Braun:2006me, Ambroso:2009jd, Anderson:2011ns, Anderson:2012yf, Ovrut:2012wg}.

On the other hand, the moduli problem of heterotic string theory poses a particular challenge due to the absence of the RR-fluxes and branes that are present in type II theories. There has been some recent progress towards understanding the heterotic moduli problem for perturbative compactifications to Minkowski vacua \cite{Anderson:2010mh, Anderson:2011ty, Anderson:2014xha, delaOssa:2014cia, GarciaFernandez:2015hja, delaOssa:2015maa}, but a full understanding is still lacking. Worse still, these compactifications always seem to be plagued with runaway directions, and it seems hard, if not impossible, to perturbatively stabilize all moduli in a maximally symmetric, large-volume scenario, and non-perturbative effects are usually required \cite{Anderson:2011cza}. This can often be attributed to the Maldacena-Nunez no-go theorem, which does not allow for fluxes in maximally symmetric Minkowski compactifications \cite{Maldacena:2000mw}. In particular, it is a simple application of Stokes' theorem to see that a closed flux cannot be used to stabilize moduli in maximally symmetric perturbative compactifications \cite{Gauntlett:2002sc}.\footnote{It should be noted that this result can be evaded by considering non-geometric compactifications in which the dilaton field is not well-defined globally, and such compactifications have recently been considered in the M-theory context \cite{Shahbazi:2015sba}.}

In recent years, a slightly different approach has become popular within the community. The idea is that instead of looking for maximally symmetric perturbative vacua, one looks for perturbative vacua with less symmetry, e.g.  domain walls \cite{lukas2011g, Klaput:2011mz, Matti:2012roa, Gray:2012md, Klaput:2013nla, de2014exploring}, thus sidestepping the no-go theorem. Moreover, the compactifications thus considered also allow for torsion on the internal space, which behaves in a similar manner to an additional flux component and hence can play a role in stabilizing moduli. The compact space is then said to have an $SU(3)$ structure, rather than $SU(3)$ holonomy, where the former is distinguished from the latter by the fact that the spinor parametrizing supersymmetry transformations is no longer covariantly constant with respect to the Levi-Civita connection.  In the domain wall case, the vacuum breaks maximal symmetry along one external coordinate direction, while the internal geometry usually corresponds to a so-called half-flat manifold.  The fibration of this internal space along the external coordinate admits a seven-dimensional $G_2$ structure.  Such vacua can subsequently be lifted to maximally symmetric ones by the inclusion of higher order $\alpha'$-effects and non-perturbative effects. This was demonstrated to work by a "proof of principle" in \cite{Klaput:2012vv}. See also \cite{Chatzistavrakidis:2008ii, Chatzistavrakidis:2009mh, Lechtenfeld:2010dr, Gemmer:2011cp, Gemmer:2012pp, Chatzistavrakidis:2012qb, Gemmer:2013ica, Haupt:2014ufa} for related work in this direction.

In this paper we will go further and generalize the perturbative compactification ansatz from the half-flat domain wall case to include a broader class of solutions; in particular, we allow a dependence of the internal geometry on up to two external coordinates. We will see that with this ansatz, the possible torsional configurations of the internal space can be more generic.  For the compact six-dimensional geometry we assume a generalization of the half-flat algebra whose fibration over the two external coordinates admits an eight-dimensional $Spin(7)$ structure.  The corresponding four-dimensional solutions are 1/4-BPS and contain the 1/2-BPS domain walls as special cases.\footnote{An alternative approach is to consider $G_2$ structures where the internal manifold is fibred over an interval that is not a Cartesian coordinate direction --- for examples of this approach, see appendix \ref{sec:csandbh}.}

Interestingly, we will see that some of the geometries give rise to superpotentials of the kind found in non-geometric type II compactifications, and thus may have interesting phenomenological applications.  However, our key goal in this paper is to derive and establish the relationships between $Spin(7)$ structures, generalized half-flat manifolds, and 1/4-BPS vacua.  Hence we defer the analysis of explicit solutions and their applications to a future publication \cite{Angus2015}. 

This paper is organized as follows.  In section \ref{sec:10d} we establish a top-down ansatz for the ten-dimensional theory.  First we review relevant aspects of ten-dimensional heterotic supergravity.  We then give a brief overview of $Spin(7)$ and $SU(3)$ structures, together with their corresponding torsion classes, and we lay the groundwork for compactification on generalized half-flat manifolds.  We also derive the corresponding Killing spinor equations. In section \ref{sec:dimred} we consider the four-dimensional theory. We derive the Killing spinor equations for four-dimensional 1/4-BPS solutions, and then we proceed to match them with the ten-dimensional ansatz. We conclude with a discussion and outlook in section \ref{sec:disc}.


\section{Ten-dimensional heterotic supergravity} \label{sec:10d}


\subsection{Killing spinor equations and metric ansatz}

The bosonic part of the ten-dimensional effective action for heterotic supergravity at lowest order in $\alpha'$ is given by
\begin{equation}\label{action10d}
S^S_{0,{\rm bosonic}}=-\frac{1}{2\kappa^2_{10}}\int_{M_{10}} e^{-2\hat\phi}\left[\hat{R}*\textbf{1}+4\d\hat\phi\wedge *\d\hat\phi-\frac{1}{2}\hat H\wedge *\hat H\right] \; ,
\end{equation}
where $\kappa_{10}$ is the 10-dimensional Planck constant, $\hat\phi$ is the dilaton, and $\hat H=d\hat B$, the field strength of the NS-NS rank-two anti-symmetric tensor field $\hat B$. Gauge field terms only arise at first order in $\alpha'$ and, therefore, do not appear in the above action. In this paper we will restrict our discussion to the lowest order in $\alpha'$ and will not consider any gauge sector fields explicitly.

The bosonic Einstein equations which follow from this action are
\begin{eqnarray}\label{Einstein}
\hat{R}_{MN}-\frac{1}{4}\hat H_{PQM}\hat H^{PQ}_{\phantom{PQ}N}+2\nabla_M\partial_N\hat\phi&=&0 \; ,\\
\nabla_M\left(e^{-2\hat\phi}\hat H^M_{\phantom{M}PQ}\right)&=&0 \; ,\\
\nabla^2\hat\phi-2\hat{G}^{MN}\partial_M\hat\phi\partial_N\hat\phi+\frac{1}{12}\hat H_{MNP}\hat H^{MNP}&=&0\; ,
\end{eqnarray}
where $\nabla$ is the covariant derivative associated with the Levi-Civita connection.

The fermionic fields associated to the bosonic fields above are the gravitino $\Psi_M$, the dilatino $\lambda$, and the gaugino $\chi$. These are all 10-dimensional Majorana-Weyl spinors with supersymmetry transformations given by
\begin{eqnarray}\label{KSE10d1}
\delta\psi_M&=&\left(\nabla_M+\frac{1}{8}{\cal\hat H}_M\right)\epsilon \; ,\\\label{KSE10d2}
\delta\lambda&=&\left(\not\!\nabla\hat\phi+\frac{1}{12}{\cal\hat H}\right)\epsilon, \; \\\label{KSE10d3}
\delta\chi&=&F_{MN}\Gamma^{MN}\epsilon \; ,
\end{eqnarray}
where $\epsilon$ is a 10-dimensional Majorana-Weyl spinor parametrizing the transformations. We also defined ${\cal\hat H}_M=\hat H_{MNP}\Gamma^N\Gamma^P$ and ${\cal\hat H}=\hat H_{MNP}\Gamma^M\Gamma^N\Gamma^P$, the contractions with gamma matrices $\Gamma^M$ satisfying the Clifford algebra in 10 dimensions.

In a supersymmetric vacuum, the above transformations must vanish, meaning that the vacuum geometry carries a covariantly constant spinor $\epsilon$ with respect to the Bismut connection,
\begin{equation}\label{Bismut}
\nabla^{(H)}_M\equiv\nabla_M+\frac{1}{8}{\cal\hat H}_M\; .
\end{equation}
Provided that the dilaton $\hat\phi$ and flux $\hat H$ satisfy~\eqref{KSE10d2}, it can be shown, neglecting the gauge fields, that solving the above Killing spinor equations leads to a solution of the equations of motion (up to first order corrections in $\alpha'$) when the Bianchi identity for $\hat H$ is satisfied. At the order we are working with, this Bianchi identity is given in its simple form as
\be
\hat H = \d\hat B \;.
\ee
We will thus focus in solving $\delta\psi_M=\delta\lambda=0$ for the remainder of this paper.

In order to recover a four-dimensional effective field theory, the background geometry must further be assumed to have six compact dimensions. Killing spinors lead to the existence of stable differential forms and, in general, such forms will reduce the structure group of the manifold they are living on. For instance, the metric of a Riemannian manifold reduces the structure group from $O(d)$ down to $SO(d)$, where $d$ is the dimension of the manifold. Considering theories preserving ${\cal N} = 1$ supersymmetry in the effective four-dimensional theory, the Killing spinors reduce the structure group of the internal six compact dimensional manifold down to $SU(3)$. For a trivial NS-NS field and a constant dilaton, this leads to the well known Calabi-Yau compactification, with the levi-Civita connection having $SU(3)$ holonomy. We can also have a non-trivial NS-NS field strength and non-constant dilaton, leading to the Bismut connection having $SU(3)$ holonomy. When the external space is Minkowski this corresponds to the Strominger solution \cite{Strominger1986253, Hull:1986kz}. In the physics literature, the terminology $SU(3)$ holonomy implicitly means that we are talking about the Levi-Civita connection, while $SU(3)$ structure refers to a more general connection having $SU(3)$ holonomy.

The aim of this work is to generalize background solutions with geometries breaking part of the four-dimensional supersymmetry, thus leading to BPS states of the effective supergravity. For this, we assume a metric ansatz with six compact internal dimensions that can depend on two of the non-compact directions, with an additional two-dimensional external Minkowski geometry:\footnote{In principle, one could allow for warp factors on the external part of the metric. However, when considering the reduction of the BPS equations on these geometries it turns out that the warp factors must be constant.}
\begin{equation}\label{ansatz10dgeneral}
\d s_{10}^2=\eta_{\alpha\beta}\d x^\alpha\d x^\beta+\underbrace{\delta_{ab}\d x^a \d x^b+\underbrace{g_{uv}(x^m)\d x^u\d x^v}_{X, \; 6d}}_{Y, \; 8d} \; ,
\end{equation}
where we let $\{m\}\in\{2,3,...,9\}$ denote coordinates on the non-compact eight-dimensional geometry, $\{u,v\}\in\{4,5,...,9\}$ denote coordinates on the compact $SU(3)$ structure manifold $X$ with metric $g_{uv}(x^m)$, $\{a,b\}\in\{2,3\}$ denote the non-compact coordinates on $Y$, which we will also denote by $x^2=x$ and $x^3=y$, while $\eta_{\alpha\beta}$ is the two-dimensional Minkowski metric with indices $\{\alpha,\beta\}\in\{0,1\}$.

This metric ansatz is a direct generalization of the usual direct (or warped) product considered for ${\cal N}=1$ solutions. However, the non-trivial dependence of the internal space on two special directions breaks some of the supercharges. In the next section, we will consider how to solve the Killing spinor equations \eqref{KSE10d1}-\eqref{KSE10d3} in this context.


\subsection{$Spin(7)$ and $SU(3)$ structures}

Let us first translate the Killing spinor equations into $Spin(7)$ geometry and further into $SU(3)$-structure forms. Under the product structure of the metric ansatz~\eqref{ansatz10dgeneral}, the eight-dimensional subspace decomposes the covariantly constant Majorana-Weyl spinor $\epsilon$ parametrizing the supersymmetry transformation according to
\begin{equation}
\begin{array}{ccc}
SO(1,9) & \supset & SO(8) \\
{\bf 16} & & {\bf 8_c} + {\bf 8_s}
\end{array} ~,
\end{equation}
where $\bf 8_c$ is the conjugate representation and $\bf 8_s$ is the spinor representation~\cite{slansky1981group}. Choosing a background geometry that breaks all but one supercharge, we will have one covariantly constant spinor preserved. This spinor will have a stability subgroup $Spin(7)$ in the $\bf 8_s$ representation, as can be seen from the branching rules
\begin{equation}
\begin{array}{ccc}\label{onesupercharge}
SO(8) & \supset & SO(7) \\
{\bf 8_s} & &  {\bf 1}+ {\bf 7}  \\
{\bf 8_c} & & {\bf 8}
\end{array} ~,
\end{equation}
the singlet ${\bf 1}$ corresponding to the unbroken supersymmetry parameter. This type of background corresponds to $1/4$-BPS states from the point of view of the effective four-dimensional ${\cal N}=1$ supergravity.

The presence of this spinor on the eight-dimensional geometry will reduce its structure group down to $Spin(7)$, the stability subgroup~\cite{Gauntlett:2003cy}. Let us recall that a $Spin(7)$ structure is given in terms of the invariant Cayley four-form $\Psi$, which may be written as
\begin{equation}
\label{4form}
\Psi_{mnpq}=\eta^\dagger\gamma_{mnpq}\eta\:,
\end{equation}
where the spinor $\eta$ denotes the supersymmetry parameter living in the $\bf 1$ of $SO(7)$. The corresponding $Spin(7)$ torsion classes are parametrized by two components, $\theta$ and $\tau$, where 
\begin{equation}
\d_8\Psi=\theta\wedge\Psi+\tau\:.
\end{equation}
Here $\theta$ and $\tau$ are in the $\mathbf 8$ and $\mathbf{48}$ representations of $Spin(7)$, respectively, while $\d_8$ is the eight-dimensional exterior derivative. See e.g.~\cite{de2014exploring} for an introduction to torsion classes and~\cite{Agricola:2006tx} for the specifics of $Spin(7)$ torsion classes.

Now, as mentioned earlier, the eight-dimensional part of the metric ansatz is further assumed to have a product structure, with two non-compact coordinates and an internal six-dimensional compact manifold. This implies the existence of two globally defined one-forms $\d x$ and $\d y$ corresponding to the non-compact directions. This defines an $SU(3)$ structure on the compact six-dimensional internal manifold, characterized by the two-form $J$ corresponding to the hermitian structure, and a complex three-form $\Omega$ corresponding to an (almost) complex structure. The $Spin(7)$ structure~\eqref{4form} decomposes under this $SU(3)$-structure embedding according to
\begin{equation}
\Psi={\rm Re}(\d z\wedge\Omega)+\frac{1}{2} J\wedge J+\d{\rm vol}_2\wedge J\;,
\end{equation}
where $\d{\rm vol}_2$ is the volume form on the non-compact directions, which by \eqref{ansatz10dgeneral} is given by
\begin{equation}
\d{\rm vol}_2=\d x\wedge\d y\:.
\end{equation}
We also have $\d z$, a complex one-form on the non-compact directions, which upon a choice of complex structure can be chosen as $\d z=\d x+i\d y$.

\subsection{Flow equations and torsion classes}

Since the structure forms are defined by the ten-dimensional covariantly constant spinor $\epsilon$, it is possible to translate the equations of motion onto properties of the structure forms. Indeed, the supersymmetry equations \eqref{KSE10d1}--\eqref{KSE10d2} can be reduced to
\begin{align}
\label{eq:susy(7)1}
*_8 \hat H=-e^{2\hat \phi}\d_8(e^{-2\hat \phi}\Psi)=&-\d_8^{\hat \phi}\Psi\;,\\
\label{eq:susy(7)2}
\theta=12\d_8\hat \phi=\Psi\lrcorner\d_8\Psi=&-*_8(\Psi\wedge*_8\d_8\Psi)\;,
\end{align}
where $*_8$ is the eight-dimensional Hodge operator and, for ease of notation, we have defined the derivative
\begin{equation}
\d_8^{\hat \phi}=e^{2\hat \phi}\d_8e^{-2\hat \phi}\;.
\end{equation}

These equations can be decomposed under the $SU(3)$ structure and, plugging the above equations into~\eqref{eq:susy(7)1}, we obtain
\begin{align}
\label{eq:flow1}
*\hat H&=\p_y^{\hat \phi}\Omega_++\p_x^{\hat \phi}\Omega_--\d^{\hat \phi} J\;,\\
\label{eq:flow2}
\d^{\hat \phi}\Omega_+&=\p_x^{\hat \phi}\rho\;,\\
\label{eq:flow3}
\d^{\hat \phi}\Omega_-&=-\p_y^{\hat \phi}\rho\;,\\
\label{eq:flow4}
\d^{\hat \phi}\rho&=0\;.
\end{align}
where $*$ is the six-dimensional Hodge operator, and $\d^{\hat \phi}=e^{2\hat \phi}\d e^{-2\hat \phi}$, with $\d$ being the usual exterior derivative in six dimensions. To simplify the notation we have also defined the forms
\begin{equation}
\rho = *J = \frac{1}{2}J\wedge J \;\;\; {\rm and} \;\;\; \Omega=\Omega_++i\Omega_-\;.
\end{equation}
Finally, from \eqref{eq:susy(7)2}, we have the conditions
\begin{align}
\label{eq:cond1}
\Omega_+\wedge \hat H&=2\,*\p_y\hat \phi\;,\\
\label{eq:cond2}
\Omega_-\wedge \hat H&=2\,*\p_x\hat \phi\;,\\
\label{eq:cond3}
 J\wedge \hat H&=-2\,*\d\hat \phi\;.
\end{align}
Solving equations~\eqref{eq:flow1}--\eqref{eq:flow4} and~\eqref{eq:cond1}--\eqref{eq:cond3} is therefore equivalent to finding solutions to the equations of motion and will be the principal focus of this work.

One question immediately arises: what kind of constraints on the geometry of the internal six-dimensional manifold follow from these flow equations? Manifolds carrying $SU(3)$ structure can be classified according to their torsion classes in the following way. The intrinsic torsion of a connection with $SU(3)$ holonomy can be classified according to five classes $\cW_i$, $i=1,\ldots,5$, with the $SU(3)$-structure forms decomposing as
\begin{equation}\label{su3torsion}
dJ=-\frac{3}{2}\,{\rm Im}\left(\cW_1\bar\Omega\right)+\cW_4\wedge J+\cW_3 \;, \quad \d\Omega=\cW_1J\wedge J+\cW_2\wedge J+\bar\cW_5\wedge\Omega \;,
\end{equation}
such that the relations
\begin{equation}
\cW_3\wedge\Omega=\cW_3\wedge J=\cW_2\wedge J\wedge J=0 \;
\end{equation}
are satisfied.

\begin{table}[ht]
\begin{center}
\begin{tabular}{|c|c|}
  \hline
  Torsion classes & Properties (name) \\
  \hline
  $\cW_1=\cW_2=0$ & Complex \\
  $\cW_1=\cW_3=\cW_4=0$ & Symplectic \\
  $\cW_1=\cW_2=\cW_3=\cW_4=0$ & K\"ahler \\
  $\cW_2=\cW_3=\cW_4=\cW_5=0$ & Nearly-K\"ahler \\
  $\cW_{1-}=\cW_{2-}=\cW_4=\cW_5=0$ & Half-flat \\
  $\cW_1=\cW_2=\cW_3=\cW_4=\cW_5=0$ & Calabi-Yau \\
  \hline
\end{tabular}
\parbox{6in}{\caption{\it\small Sample of six-dimensional $SU(3)$-structure properties determined in terms of their vanishing torsion classes. For the half-flat case, the subscript $-$ means the imaginary part.}\label{tablesu3torsion}}
\end{center}
\end{table}

Some geometrical properties can readily be obtained from these torsion classes, depending on whether they vanish or not --- some well-known cases are summarized in table~\ref{tablesu3torsion}. Therefore, we wish to derive the internal torsion classes $\cW_i$ in terms of the field content of the theory to see what kinds of restrictions the flow equations impose, in particular in terms of vanishing classes.

First we define the complexified coordinate
\be\label{z}
z=x+iy\;.
\ee
Thus we can combine and rewrite equations~\eqref{eq:flow2} and~\eqref{eq:flow3} as
\begin{equation}
\label{eq:dOmega}
\d\Omega-2\d\hat \phi\wedge\Omega=2\p_z\rho-4\p_z\hat\phi\rho=2\p_zJ\wedge J-2\p_z\hat \phi\,J\wedge J\;.
\end{equation}
We see that $\p_z\hat \phi$ contributes to $\cW_1$ and $\p_zJ$ will contribute to $\cW_2$ and $\cW_5$. Since $J$ together with $\Omega$ define an (almost) complex structure, with respect to which $\Omega$ is a $(3,0)$-form, we can decompose $\p_zJ$ according the holomorphicity of its indices. We define
\begin{equation}
\p_zJ=\lambda_zJ+h^{(1,1)}_z+h_z^{(2,0)+(0,2)}\;,
\end{equation}
where $\lambda_z$ is a zero-form and $h^{(1,1)}_z$ represents the primitive $(1,1)$-part, that is, it must satisfy the condition $J\,\lrcorner \,h^{(1,1)}_z=0$. Equation~\eqref{eq:dOmega} can then be written as
\begin{equation}
\d\Omega=(\lambda_z-2\p_z\hat \phi)J^2+\left(h_z^{(1,1)}+h_z^{(2,0)}\right)\wedge J+2\d\hat \phi\wedge\Omega\; ,
\end{equation}
where we used the fact that, since $\d\Omega$ cannot be $(1,3)$, we must have $h_z^{(0,2)}=0$. We can therefore deduce the first two torsion classes,
\begin{equation}
\cW_1=\lambda_z-2\p_z\hat \phi\;,\;\;\;\cW_2=h_z^{(1,1)}\;.
\end{equation}

In order to find $\cW_5$, we must understand the contribution that comes from the $h_z^{(2,0)}$ term. Let us write
\begin{equation}
h_z^{(2,0)}\wedge J=X\wedge\Omega\;,
\end{equation}
where $X$ is a $(0,1)$-form and is to be determined. Using formulae of the appendix of \cite{lukas2011g}, and $\vert\vert\Omega\vert\vert^2=8$ as is conventional, we find that
\begin{equation}
\cW_5=2(\d\hat \phi)^{(1,0)}+\frac{i}{8}\left(h_z^{(2,0)}\lrcorner\Omega\right)\;.
\end{equation}

The final two torsion classes can readily be determined from the remaining conditions. We see from~\eqref{eq:flow4} that
\begin{equation}\label{W4}
\cW_4=\d\hat \phi\;
\end{equation}
and, using~\eqref{eq:flow1} and subtracting the contributions from $\cW_3$ and $\cW_4$, we find that the last torsion class is
\begin{equation}
\cW_3=\p_y\Omega_+-2\p_y\hat \phi\,\Omega_++\p_x\Omega_--2\p_x\hat \phi\,\Omega_-+\frac{3}{2}\textrm{Im}\left(\cW_1\bar\Omega\right)+\d\hat \phi\wedge J-*\hat H\;.
\end{equation}

To conclude, we have seen that the flow equations impose constraints on the torsion classes; however, they do not impose any vanishing classes. Therefore, they constitute a generalization of the half-flat compactifications obtained from half-BPS flows~\cite{lukas2011g}.

\subsection{Internal geometry ansatz}

We have seen that the existence of an $SU(3)$ structure implies that the compact internal six-dimensional manifold carries a two-form $J$ and a three-form $\Omega$. In order to consider the effective four-dimensional theory resulting from compactification, one must have an understanding of the moduli space of metrics.

In general, explicit examples of $SU(3)$ structure manifolds are scarce, so we would like to take a more general approach to deal with the Kaluza-Klein truncation to four dimensions. For this, we draw on the results from mirror symmetry and the subsequent development concerning the so-called half-flat mirror manifolds~\cite{deCarlos:2005kh}. We expand the $SU(3)$-invariant forms on a set of two-forms $\{\omega_i\}$ and three-forms $\{\ax_A,\bx^A\}$,
\begin{equation}\label{expansion}
J=v^i\omega_i\; ,\;\;\;\Omega={\cal Z}^A\alpha_A-\mathcal{G}_A\beta^A\; ,
\end{equation}
and assume, in complete analogy with the Calabi-Yau case, that the basis forms satisfy the normalization integrals,
\begin{equation}
\int \omega_i\wedge\tilde\omega^j=\delta_i^j\; , \;\;\; \int \alpha_{A}\wedge\alpha_{B}=0\; , \;\;\; \int \beta^{A}\wedge\beta^{B}=0\; , \;\;\; \int \alpha_{A}\wedge\beta^{B}=\delta_{A}^{B}\; .
\end{equation}
where $\{\tilde\omega^j\}$ is a set of four-forms dual to $\{\omega_i\}$.

When the $SU(3)$ structure reduces to a Calabi-Yau manifold, that is, the structure is invariant with respect to the Levi-Civita connection, the form $J$ corresponds to the K\"ahler form and $\Omega$ to the holomorphic $(3,0)$ form. In this instance, $\{\omega_i\}$ forms a basis of harmonic $(1,1)$ forms of the second cohomology group while $\{\ax_A,\bx^A\}$ is a real symplectic basis of the third cohomology group. The moduli space is characterized by the deformations of $J$ and $\Omega$, where $v^i$ are the K\"ahler moduli and $Z^A$ are projective coordinates on the complex structure moduli. $\mathcal{G}=\mathcal{G}(Z^A)$ is a holomorphic function of the projective coordinates, which is homogeneous of degree two, while $\mathcal{G}_A=\p_A\mathcal{G}$.

For more general $SU(3)$-structure manifolds, we will assume that the basis forms $\{\omega_i\}$ and $\{\ax_A,\bx^A\}$ satisfy the same properties as for Calabi-Yau manifolds, with the only exception that they no longer need to be closed. This postulate arose originally in the context of mirror symmetry, where the matching of the space of metrics implies that the expansion of forms should have similar properties. Half-flat mirror manifolds have been proposed in~\cite{Gurrieri:2004dt} and generalized in~\cite{Gurrieri:2007jg, d2005gauging}. This class of spaces has been argued to be the correct ansatz for a consistent Kaluza-Klein truncation to four dimensions~\cite{tomasiello2005topological,grana2006hitchin}.

Following the conventions of~\cite{d2005gauging,grana2006hitchin,deCarlos:2005kh}, we consider the so-called \emph{generalized half-flat manifolds}, whose $SU(3)$-structure forms $\omega_i$ and $(\ax_A,\bx^B)$ are postulated to obey the 
algebra
\begin{equation}
 \d \ox_i = p_{Ai} \bx^A - q^A_i \ax_A \; ,\;\;\;
 \d \ax_A  = p_{Ai} \tox^i\; ,\;\;\;  \d \bx^A  = q^A_i \tilde{\omega}^i
 \; ,\;\;\; \d \tox^i=0\; ,\label{extalg}
\end{equation}
with (real) constant torsion parameters $p_{Ai}$ and $q_i^A$. These are the most general expressions, assuming that the exterior derivatives $\d J$ and $\d\Omega$ can be expanded on the basis of three-forms $\{\ax_A,\bx^A\}$ and four-forms $\{\tilde\omega^j\}$, respectively. Explicitly, this gives
\begin{equation}
\d J=v^ip_{Ai} \bx^A - v^jq^A_j \ax_A\; ,\;\;\; \d\Omega={\cal Z}^Ap_{Ai}\,\tilde\omega^i-\mathcal{G}_Aq^A_j\,\tilde\omega^j\; . \label{dJdOmega}
\end{equation}
For consistency, we still need to impose additional constraints on the parameters $p_{Ai}$ and $q_i^A$ in order to satisfy the condition $\d^2\omega_i=0$. The flux parameters must obey the relations
\begin{equation}
  \label{cons1}
  p_{Ai} q^A_j - q^A_i p_{Aj} = 0 \; 
\end{equation} 
for a consistent definition of the exterior derivative. In terms of torsion classes, the definitions~\eqref{dJdOmega} imply that the intrinsic torsion $\tau^0$ must take value in the modules
\begin{equation}
  \tau^0\in\cW_1\oplus\cW_2\oplus\cW_3 \;.
\end{equation}
It should be noted that this particular ansatz imposes $\cW_4=\cW_5=0$. This will have some consequences on the solutions of the flow equations, as explained in the next section.

Special cases can readily be found in this general context. First, half-flat mirror manifolds correspond to the particular choice
$p_{0i}=e_i$, $p_{ai}=0$, and $q_i^A=0$, with the basis forms obeying
\begin{equation}
\d\omega_i=e_i\beta^0\; ,\quad \d\alpha_0=e_i\tilde{\omega}^i\; , \label{hfmdef}
\end{equation}
and being closed otherwise. This leads to
\begin{equation}
\d J=v^ie_i\beta^0\; ,\quad \d\Omega={\cal Z}^0e_i\tilde\omega^i\; , \label{dJ}
\end{equation}
which satisfy the definition of half-flat manifolds.  For the intrinsic torsion, we have
\begin{equation}
  \tau^0\in\cW_1^+\oplus\cW_2^+\oplus\cW_3 \;.
\end{equation}
Second, when all the torsion parameters vanish, that is $p_{Ai}=q^A_j = q^A_i= p_{Aj} = 0$, we fully recover the Calabi-Yau geometry with the relations $\d J=\d\Omega=0$.  Hence in this case
\begin{equation}
  \tau^0=0 \;.
\end{equation}

This ansatz for the basis forms also has an impact on the NS-NS flux that can be considered for the compactification. Owing to the generalized relations for the exterior derivatives~\eqref{extalg}, expanding the NS-NS three-form field strength onto the truncation basis gives
\begin{equation}
 \hat{B}=B+b^i\omega_i\; , \quad\hat{H}=\d_{10}\hat{B} = H+\d_4 b^i\wedge\omega_i+b^i\d\omega_i+H_{\rm
  flux}\label{Bzero} \; ,
\end{equation}
where the $b^i$ are axionic scalars which we take to be internally constant, and $B$ is a four-dimensional two-form with field strength $H=\d_4 B$, which can be dualized to a universal axion $a$. In addition, we have introduced the NS-NS flux,
\begin{equation}
  \label{Hflux1}
  H_{\textrm{flux}} = \mu^A \ax_A - \ex_A \bx^A \; ,
\end{equation}
with electric and magnetic flux parameters $\ex_A$ and $\mu^A$, respectively. Constraints on the electric and magnetic flux parameters occur according to the choice of gauge bundle. Choosing the order $\alpha'$ term to vanish in the Bianchi identity implies $\d\hat{H} = 0$ and leads to the relations
\begin{equation}
  \label{cons2}
  \mu^A p_{Ai} - \ex_A q^A_i = 0 \; 
\end{equation}
between flux and torsion parameters. As a simplifying assumption, we will only consider solutions where the flux has legs only on the internal space, that is,
\begin{equation}\label{eq:const-b-a}
H_{MNa}=H_{MN\alpha}=0\;.
\end{equation}
Note from \eqref{Bzero} that this assumption implies that we need to take the axions $\{b^i,a\}$ constant when performing the matching with the four-dimensional theory.

Finally, we should consider how this ansatz constrains the flow equations.  In particular, for our specific choice of internal metric, we have $\cW_4=0$.  Comparing with~\eqref{W4}, we see that the six-dimensional exterior derivative of the dilaton must vanish, $\d\hat \phi=0$. This brings the following simplification of the flow equations~\eqref{eq:flow1}--\eqref{eq:flow3}:
\begin{align}
\label{eq:newflow1}
\d J&=2\textrm{Im}\left(\p_{\bar z}\Omega-2\p_{\bar z}\hat \phi\Omega\right)-*\hat H \; ; \\
\label{eq:newflow2}
\d\Omega&=2\partial_z\rho -4\partial_z\hat \phi \rho \; .
\end{align}
From the basis form properties we have $\d\rho=0$, so it follows that equation~\eqref{eq:flow4} is satisfied automatically.
Furthermore, the conditions~\eqref{eq:cond1}--\eqref{eq:cond2} become
\be
\label{eq:newcond}
\Omega\wedge \hat H=4i\,*\p_z\hat \phi\;,
\ee
Since we have $\d\hat \phi=0$, it follows from~\eqref{eq:cond3} that $J\wedge \hat H = 0$, and hence this third condition corresponds to
\be \label{eq:bi}
\d b^i=0\:,
\ee
which is true since the axions are taken to be constant internally.

The equations \eqref{eq:newflow1}--\eqref{eq:newcond} will be our starting point in section \ref{sec:match} when we match to the four-dimensional flow equations, to which we now turn.


\section{Dimensional reduction}
\label{sec:dimred}

\subsection{Four-dimensional effective theory}
\label{sec:4d}

Having established our conventions for the internal compact dimensions, we can work out the resulting effective four-dimensional theory that follows from Kaluza-Klein reduction. For this, we must assume that the flux parameters are small enough compared to the volume of the internal manifold to allow a separation between the string scale and the flux scale. The relations defined in the previous section are then sufficient to derive the low-energy superpotential of the effective supergravity theory.

Considering the ten-dimensional action and integrating out the compact dimensions, we assume a field truncation that preserves the expansion of the modes on the basis forms $\{\omega_i,\ax_A,\bx^A\}$. In terms of the internal volume, $\mathcal{V} = \int d^6x \sqrt{g}$, the four-dimensional dilaton is given by
\begin{equation}\label{eq:dilaton}
\phi=\hat{\phi}-\frac{1}{2}{\rm ln}\mathcal{V} \; ,
\end{equation}
where $\mathcal{V}$ is the zero-mode of the ten-dimensional dilaton.

Combined with the K\"ahler and complex structure moduli~\eqref{expansion}, the various flux scalar fields form the lowest components of the four-dimensional chiral supermultiplets in the usual way,
\begin{equation}
S=a+ie^{-2\phi}\; ,\;\;\; T^i=b^i+iv^i\;, \;\;\; Z^a=z^a \equiv c^a +i \omega^a\; . \label{superfields}
\end{equation}
The corresponding K\"ahler potentials are given by the same expressions as for Calabi-Yau compactifications, coming from the logarithm of the internal volume,\footnote{For useful background and formulae, see appendix \ref{moduli}.}
\begin{equation}
K^{(S)}=-\ln(i(\bar{S}-S))\; ,\quad K^{(T)}=-\ln(\frac{4}{3}\int J \wedge J \wedge J)\; ,\quad K^{(Z)}=-\ln\left(i\int\Omega\wedge\bar{\Omega}\right)\; .
\label{hfk}
\end{equation} 
The full K\"ahler potential of the resulting four-dimensional effective supergravity thus corresponds to the sum of the three terms above,
\begin{equation}
K = K^{(S)} + K^{(T)} + K^{(Z)}\; .
\label{K}
\end{equation} 

The superpotential can be obtained from the Gukov-Vafa formula, as derived in~\cite{deCarlos:2005kh},
\begin{equation}
W=\sqrt{8}\int\Omega\wedge(\hat H+i\d J) \;. \label{W}
\end{equation} 
For convenience, let us introduce the modified flux parameters,
\begin{equation}
\label{cpxflux}
\begin{aligned}
\te_A & = \ex_A - T^i p_{Ai} \;, \\
\tm^A & = \mu^A -T^i q^A_i \; .
\end{aligned}
\end{equation}
These allow us to express the exterior derivative of the hermitian form $\d J$ and the NS-NS field strength $\hat{H}$ in the compact form
\begin{equation}
\label{Hint}
\d J = v^i \d\ox_i = \im (\tm^A)\ax_A -\im (\te_A)\bx^A \; , \;\;\; \hat{H} = 
\re (\tm^A) \ax_A - \re (\te_A) \bx^A \; ,
\end{equation}
where we have imposed \eqref{eq:const-b-a} and \eqref{eq:bi}.  This in turn leads to the superpotential expression
\begin{equation}
W = \sqrt{8} \ (\tm^A\cG_A - \te_A \cZ^A) \;. \label{eq:W_moduli}
\end{equation}

It should be remembered that, owing to the dependence on the complexified K\"ahler moduli $T^i$, the new parameters $\te_A$ and $\tm^A$ can no longer be taken to be constant. It is interesting to note that superpotentials of type \eqref{eq:W_moduli} are very similar to those that appear in type~II non-geometric compactifications \cite{shelton2005nongeometric, aldazabal2006more, Grana:2006hr, micu2007towards}. These types of superpotentials have recently been employed in order to obtain phenomenologically viable models of Starobinsky-like inflation \cite{hassler2014inflation, blumenhagen2015towards}. It would be interesting to see if such inflationary models can be obtained within the heterotic context, and also if other phenomenological applications apply. We leave this for future work.

\subsection{Killing spinor equations and $1/4$-BPS states}
\label{sec:14BPS}

The four-dimensional effective theory resulting from heterotic string theory is an $\cN=1$ \hbox{supergravity}. In general, such a theory is characterized by the chiral superfields $(A^I, \chi^I)$, a gravitino $\psi_\mu$, the K\"ahler potential $K$, and the superpotential $W$. The corresponding Killing spinor equations are given by
\begin{eqnarray}\label{eq:KSE4d1}
\delta\chi^I&=&i \sqrt{2} \sigma^\mu \bar{\zeta} \partial_\mu A^I-\sqrt{2}e^{K/2}K^{IJ^*}D_{J^*}W^*\zeta=0\; ,\\\label{eq:KSE4d2}
\delta\psi_\mu&=&2 \mathcal{D}_\mu\zeta+ie^{K/2}W\sigma_\mu\bar{\zeta}=0\; ,
\end{eqnarray}
where the covariant derivative is defined by
\begin{equation}
\mathcal{D}_\mu=\partial_\mu+\omega_\mu+\frac{1}{4}(K_j\partial_\mu A^j-K_{j^*}\partial_\mu A^{j^*})\; .
\end{equation} 
In the above $\zeta$ is a Weyl spinor that parametrizes supersymmetry, \hbox{$D_IW=W_I+K_IW$,} the $\sigma_\mu$~correspond to the usual Pauli matrices, and $\omega_\mu$ is the spin connection.

In order to match with the eight-dimensional $Spin(7)$ structure, we should focus in the four-dimensional theory on topological defects of co-dimension two.  With this in mind, we consider the metric ansatz
\begin{equation}
 ds_4^2=e^{-2B(x^a)}\left(\eta_{\alpha\beta}\d x^\alpha \d x^\beta +\delta_{ab}\d x^a \d x^b\right) \label{eq:4dmetric} \; ,
\end{equation}
where $\{\alpha, \beta\} = \{0,1\}$, while the $\{a,b\} = \{2,3\}$ directions parametrize the $xy$-plane.
From this metric ansatz one obtains the spin connection
\begin{equation}
\omega_0 = \frac{1}{2}B_a\sigma^{\underline{a}} \; , \;\;\;
\omega_1 = -i\frac{1}{2}\left(B_2\sigma^{\underline{3}} - B_3\sigma^{\underline{2}}\right) \; , \;\;\;
\omega_2 = -i\frac{1}{2}B_3\sigma^{\underline{1}} \; , \;\;\;
\omega_3 = i\frac{1}{2}B_2\sigma^{\underline{1}} \; ,
\end{equation}
where $B_a \equiv \partial B/\partial x^a$.
Applying this result to (\ref{eq:KSE4d1}) and (\ref{eq:KSE4d2}) gives
\begin{align}
A^I_a\sigma^{\underline{a}} \overline{\zeta} &= -ie^{-B}e^{K/2}K^{IJ^*}D_{J*}W^* \zeta \; , \label{eq:KSEaA} \\
B_b\sigma^{\underline{b}} \zeta &= ie^{-B}e^{K/2}W \overline{\zeta} \; , \label{eq:KSEbB} \\
0 &= 2\zeta_a + \left(B_a + i\text{Im}(K_IA^I_a)\right)\zeta \; . \label{eq:KSEaB}
\end{align}
Since the warp factor $B$ is real, we can separate (\ref{eq:KSEaB}) into real and imaginary parts,
\begin{eqnarray} \label{eq:KSE3}
\text{Im}(K_IA^I_a) &=& 0 \; , \\
2\zeta_a &=& -B_a\zeta \; . \label{eq:KSE4}
\end{eqnarray}

We are interested in solutions preserving only one of the four supercharges from $\zeta$, which corresponds to the singlet of~\eqref{onesupercharge}. To this end, we impose on the Killing spinor the constraints
\begin{equation}
\sigma^{\underline{2}}\, \zeta = \overline{\zeta} \; , \;\;\;
i \sigma^{\underline{3}}\, \zeta = \overline{\zeta} \; , \label{eq:1-4BPS}
\end{equation}
effectively breaking three out of the four supercharges.\footnote{Note that we could instead have chosen a constraint of the form $\sigma^{\underline{2}}\, \zeta = -i \sigma^{\underline{3}}\, \zeta = \overline{\zeta}$ (or any equivalent permutation of $\sigma$-matrices) while retaining consistency of the equations.  However, this choice does not give the correct matching to our parametrization of the $Spin(7)$ geometry in section \ref{sec:10d}.  It is also possible to include an arbitrary relative phase between $\sigma^{\underline{2}}\,\zeta$ and $\bar\zeta$, but this can always be absorbed into a redefinition of the superpotential.}
Solutions will thus correspond to $1/4$-BPS supergravity states. Plugging this into the equations~\eqref{eq:KSEaA},~\eqref{eq:KSEbB}, and~\eqref{eq:KSE3} we obtain
\begin{align}
(\partial_x + i \partial_y)\, A^I &= -ie^{-B}e^{K/2}K^{IJ^*}D_{J*}W^* \; , \nn \\
(\partial_x - i \partial_y)\, B &= ie^{-B}e^{K/2}W \; , \nn \\
0 &= \text{Im}(K_I\partial_xA^I) \; , \nn \\
0 &= \text{Im}(K_I\partial_yA^I) \; . \label{eq:KSE2xDW}
\end{align} 
The last equation~\eqref{eq:KSE4} simply gives the dependence of the spinor $\zeta$ on the special directions $\{x,y\}$ as a function of the warp factor dependence $B$. We will thus ignore this condition as it can always be solved appropriately.

We can now insert the heterotic supergravity ingredients~\eqref{superfields},~\eqref{hfk}, and~\eqref{W} into these Killing spinor equations~\eqref{eq:KSE2xDW}, and using the complex coordinate $\bar z$ we obtain
\begin{align}
2\p_{\bar z}\, A^I &= -ie^{-B}e^{K/2}K^{IJ^*}D_{J*}W^* \; , \label{eq:KSEA} \\
2\p_{\bar z}\, B &= -ie^{-B}e^{K/2}W^* \; , \label{eq:KSEB} \\
0&=3\,\frac{\mathcal{K}_i}{\mathcal{K}}\p_z\,b^i+3\,\frac{\tilde{\mathcal{K}}_a}{\tilde{\mathcal{K}}}\p_z\,c^a+e^{2\phi}\p_z\,a\; . \label{eq:KSEaxion}
\end{align}

To round off this discussion, note that \eqref{eq:KSEaA}--\eqref{eq:KSE4} are invariant under general coordinate transformations in the $xy$-plane.  This suggests that there may exist \hbox{1/2-BPS} solutions other than domain wall solutions.  We explore this and related ideas further in appendix \ref{sec:csandbh}.

\subsection{Matching the $10d$ and $4d$ flow equations}
\label{sec:match}

\subsubsection*{Dilaton}
We will now try to match these flow equations with those obtained from the ten-dimensional perspective. First, let us look at the equation for the axion-dilaton field.
From~\eqref{eq:KSEA} we have
\be
2\p_{\bar z}\, (a+ie^{-2\phi}) = -ie^{-B}e^{K/2}(-2ie^{-2\phi})W^*\;.
\ee
Remembering that we made the ansatz~\eqref{eq:const-b-a} for the NS-NS flux $\hat H$, the corresponding four-dimensional axion $a$ is constant. Hence inserting~\eqref{eq:KSEB} into the above equation gives
\be
i\p_{\bar z}e^{-2\phi} =-2ie^{-2\phi}\p_{\bar z}B \; ,
\ee
which implies
\be
\p_{\bar z}\phi = \p_{\bar z}B \;,
\ee
and therefore $\phi - \phi_0 = B$ for constant $\phi_0$. From now on, we will fix the warp factor to be equal to the four-dimensional dilaton, $\phi = B$.
The flow equation for the dilaton is therefore given by
\be
2\,\p_{\bar z} \phi = -ie^{-\phi}e^{K/2}W^* \; .
\ee

We would now like to derive this from the 10d flow equations.  First take the wedge product of \eqref{eq:newflow1} with $\bar\Omega$, giving
\bea
\left(\d J + *\hat H\right)\wedge\bar\Omega
&=&-i\left(\left(\p_{\bar z}\Omega-2\,\p_{\bar z}\hat\phi\,\Omega\right)-\left(\p_{z}\bar\Omega-2\,\p_{ z}\hat\phi\,\bar\Omega\right)\right) \wedge\bar\Omega \nn \\
&=& -i\left(\p_{\bar z}\Omega-2\,\p_{\bar z}\hat\phi\,\Omega\right) \wedge\bar\Omega
\; , \label{eq:dilaton-omega}
\eea
where the second line follows from the fact that $\Omega$ is a (3,0)-form. Now, integrate \eqref{eq:newcond} over the compactification volume.  Remembering that the volume $\cV$ of the internal manifold can be written as $8\cV = i \int\Omega\wedge\bar\Omega$, we obtain
\be
\int \Omega\wedge\bar\Omega\,\p_z\hat\phi 
= -2\int \Omega\wedge \hat H
= -2i\int *\Omega\wedge\hat H = -2i\int *\hat H\wedge\Omega \; ,
\ee
where in the second step we have used the property $*\Omega = -i\Omega$.
Taking the complex conjugate gives
\be
2\int *\hat H\wedge\bar\Omega = i\int\Omega\wedge\bar\Omega\,\p_{\bar z}\hat\phi \; . \label{eq:dilaton-cons}
\ee
Integrating over \eqref{eq:dilaton-omega} and inserting \eqref{eq:dilaton-cons} on the right-hand side, we find
\be
\int\left(\d J\wedge\bar\Omega + *\hat H\wedge\bar\Omega\right) = -i\int \p_{\bar z}\Omega\wedge\bar\Omega + i\,\p_{\bar z}\hat\phi\int\Omega\wedge\bar\Omega + 2\int *\hat H\wedge\bar\Omega \; .
\ee
Rearranging and simplifying leads to
\be
-i\int \bar\Omega\wedge\left(\hat H - i\,\d J\right) = -i\int\p_{\bar z}\Omega\wedge\bar\Omega + i\,\p_{\bar z}\hat\phi\int \Omega\wedge\bar\Omega \; . \label{eq:dilaton-complex}
\ee

In order to complete the matching we will need to make use of an additional constraint, 
\begin{equation}
\int\p_z\Omega\wedge\bar\Omega = \int\Omega\wedge\p_z\bar\Omega \; . \label{eq:extra10dconstraint}
\end{equation}
This might seem like an extra condition imposed on the ten-dimensional geometry. We now show that we are free to impose this condition, at least as far as physics is concerned. Indeed, recall by Kodaira \cite{kodaira1958deformations} that
\begin{equation}
\label{eq:GenKodaira}
\p_z\Omega=K_z\Omega+\chi_z\;,
\end{equation}
where $K_z$ is some complex function and $\chi_z\in\Omega^{(2,1)}(X)$. The absolute value of $K_z$ is related to the change in volume. We are however free to rotate $K_z$ by a complex phase without changing the volume or the complex structure of $X$. In particular, we take $K_z$ to be real,
\begin{equation}
\label{eq:ChosseK}
K_z=\bar K_z\:.
\end{equation}
However,
\begin{equation}
\p_z\bar\Omega=\bar K_z\bar\Omega+\bar\chi_z\:.
\end{equation}
Hence with the choice \eqref{eq:ChosseK}, the result \eqref{eq:extra10dconstraint} follows.

Applying this condition, we see that \eqref{eq:dilaton-complex} can be expressed as
\be
-i\int \bar\Omega\wedge\left(\hat H - i\,\d J\right) = i\,\p_{\bar z}\hat\phi\int \Omega\wedge\bar\Omega - \frac{i}{2}\int\left(\p_{\bar z}\Omega\wedge\bar\Omega + \Omega\wedge\p_{\bar z}\bar\Omega\right) \; .
\ee
Inserting the definitions of the superpotential \eqref{W} and the volume $\cV$, we find
\be
-\frac{iW^*}{\sqrt{8}}
= 8\cV\,\p_{\bar z}\left(\hat\phi - \frac{1}{2}{\rm ln}\cV\right) \; .
\ee
Finally, using $e^{-\phi}e^{K/2} = 1/(4\sqrt{8}\cV)$ (derived in appendix \ref{moduli})\footnote{This condition is the same as equation \eqref{eq:K4OV} under the convention $\vert\vert\Omega\vert\vert^2=8$.} and the relation \eqref{eq:dilaton} between the 10d and 4d dilaton, we arrive at the flow equation
\begin{equation}
2\,\p_{\bar z}\phi = -ie^{-\phi}e^{K/2}W^* \; . \label{eq:dilatoncomplete}
\end{equation}

\subsubsection*{Complex structure moduli}
We now turn to the complex structure moduli.  The flow equation we wish to derive is
\begin{equation}
2\,\p_{\bar z}Z^a = -ie^{-B}e^{K/2}K^{ab^*}D_{b*}W^* \; . \label{eq:cplxstr}
\end{equation}
Here we will make use of Kodaira's formula \eqref{eq:GenKodaira}, which we assume can be written as\footnote{\eqref{eq:Kodaira} is true in the Calabi-Yau case, but has not been shown to be true for the most generic geometries that we consider. However, since the generalized half-flat manifolds are expected to be mirror to Calabi-Yau's with flux, we regard \eqref{eq:Kodaira} as a valid assumption.}
\begin{equation} \label{eq:Kodaira}
\frac{\p\Omega}{\p Z^a} = -\frac{\p K^{\rm (Z)}}{\p Z^a}\Omega + \chi_a \; ,
\end{equation}
from which it follows that the K\"{a}hler metric for the complex structure moduli can be expressed as
\begin{equation}
K^{\rm (Z)}_{ab^*} = -\frac{\int\chi_a\wedge\bar{\chi_b}}{\int\Omega\wedge\bar{\Omega}} \; . \label{eq:Kab-Z}
\end{equation}

To proceed, we start from the 10d equation \eqref{eq:newflow1}, then wedge with $\bar\chi_a$ and integrate to obtain
\be
\int\bar\chi_a\wedge (\d J+*\hat{H}) = \int\bar\chi_a\wedge \left(2\,\im\left(\p_{\bar z}\Omega - 2\p_{\bar z}\hat{\phi}\,\Omega\right)\right) 
= -i\int\bar\chi_a\wedge\p_{\bar z}\Omega \; , \label{eq:cplxstr-omega}
\ee
where in the second step we have used the fact that $\bar\chi_a$ is a (1,2)-form.
We can use Kodaira's formula to rewrite this as
\be
-i\int \bar\chi_a\wedge\p_{\bar z} \Omega
=  -i\int\bar\chi_a\wedge \chi_b\,\p_{\bar z}\, Z^b + i\int \underbrace{\bar\chi_a\wedge\Omega}_{=0}K_b\,\p_{\bar z}Z^b 
= -i\int\bar\chi_a\wedge\chi_b\,\p_{\bar z} Z^b \; .
\ee
where the second term again vanishes by the antisymmetry of the wedge product, using that $\bar\chi_a$ is type $(1,2)$ while $\Omega$ is type $(3,0)$.
Inserting this expression into \eqref{eq:cplxstr-omega} and using $*\chi = i\chi$ gives
\be
\int\bar\chi_a\wedge\chi_b\,\p_{\bar z} Z^b = i\int\bar\chi_a\wedge (\d J+*\hat{H})
= -\int\bar\chi_a\wedge \left(\hat{H}-i\,\d J\right) \; .
\ee

We are now in a position to obtain the four-dimensional flow equation.
Dividing by the volume $\cV$ and using $i\int\Omega\wedge\bar\Omega = 8\cV$, we get
\be
\,\frac{\int\bar\chi_a\wedge\chi_b}{\int\Omega\wedge\bar\Omega}\,\p_{\bar z}Z^b = -\frac{i}{8\cV}\int\bar\chi_a\wedge(\hat{H}-idJ) \; ,
\ee
and by substituting \eqref{eq:Kab-Z} we find
\be
2K_{a^*b}\,\p_{\bar z}Z^b = -ie^{-\phi}e^{K/2}\sqrt{8}\int D_{a^*}\bar\Omega\wedge(\hat{H}-idJ) \; ,
\ee
where we have again used $e^{-\phi}e^{K/2} = 1/(4\sqrt{8}\cV)$.  Note that $\bar\chi_a = D_{a^*}\bar\Omega$ follows directly from Kodaira's formula.  Finally, contracting with the inverse K\"{a}hler metric leads to the result,
\begin{equation}
2\,\p_{\bar z}Z^a = -ie^{-\phi}e^{K/2}K^{ab^*}D_{b^*}W^* \; .
\end{equation}

\subsubsection*{K\"{a}hler moduli}

Finally, let us consider the K\"{a}hler moduli.  Our goal is the 4d flow equation
\begin{equation}
2\,\p_{\bar z}T^i = -ie^{-\phi}e^{K/2}K^{ij^*}D_{j^*}W^* \; . \label{eq:T-goal}
\end{equation}

From the 10d perspective we start from \eqref{eq:newflow2}, which can be written as
\be
\d\Omega
= \p_z\left( J\wedge J\right) - 2\,\p_z \hat{\phi}\, J\wedge J \; .
\ee
Taking the wedge product with the basis form $\omega_i$, and integrating over the manifold, gives
\be
\int\d\Omega\wedge\omega_i = \int\left(\p_z\left( J\wedge J\right) - 2\,\p_z \hat{\phi}\, J\wedge J\right)\wedge\omega_i \; .
\ee
Since our final result contains only $\p_{\bar z}$ terms, we should take the complex conjugate of this equation.  After integrating by parts, we find
\be
\int\bar\Omega\wedge\d\omega_i = \p_{\bar z} \int J\wedge J\wedge\omega_i - 2\,\p_{\bar z} \hat{\phi}\int J\wedge J\wedge\omega_i \; . \label{eq:Kahler-J}
\ee

In what follows we will make use of the formulae given in appendix \ref{moduli}. To proceed, note that the superpotential \eqref{W} can be expressed in the form
\begin{equation}
W = \sqrt{8}\int\Omega\wedge\left(H_\text{flux} + \d\left(T^i\omega_i\right)\right) \; . \label{eq:WTexplicit}
\end{equation}
From \eqref{eq:WTexplicit} and the definitions \eqref{metric1}, we see that \eqref{eq:Kahler-J} can be rewritten as
\be
\p_{\bar z} \cK_i - 2\,\p_{\bar z} \hat{\phi}\, \cK_i = \frac{1}{\sqrt{8}}\frac{\p W^*}{\p \bar{T}^i} \; .
\ee
Inserting the relation \eqref{eq:dilaton} between the 10d and 4d dilaton, with $\cK = 6\cV$, we get
\be
\p_{\bar z} \cK_i - \cK_i \p_{\bar z}\text{ln}\cK = \frac{1}{\sqrt{8}}\frac{\p W^*}{\p \bar{T}^i} + 2\,\p_{\bar z} \phi\,\cK_i \; . \label{eq:Kln}
\ee

To complete the matching, we should write this in terms of the moduli $v^i$.  From \eqref{metric1} we can derive
\begin{equation}
\p_{\bar z}\cK = 3\cK_i\p_{\bar z}v^i \; , \;\;\; \p_{\bar z}\cK_i = 2\cK_ij\p_{\bar z}v^j \; . \label{eq:Kinv}
\end{equation}
Dividing \eqref{eq:Kln} by $\cK$, rescaling, and simplifying the result using \eqref{eq:dilatoncomplete}, \eqref{eq:Kinv}, \eqref{eq:Ki_contract}, and the relation \hbox{$e^{-\phi}e^{K/2} = 1/(4\sqrt{8}\cV)$,} we obtain
\be
2\left(\frac{9}{4}\frac{\cK_i\cK_j}{\cK^2} - \frac{3}{2}\frac{\cK_{ij}}{\cK}\right)\p_{\bar z}v^j = -e^{-\phi}e^{K/2}\left(\frac{\p W^*}{\p \bar{T}^i} + K_{i^*}W^*\right) \; . \label{eq:KinvDW}
\ee
The term in brackets on the left-hand side is simply the K\"{a}hler metric \eqref{eq:Kij_inverse}, while the right-hand side contains the covariant derivative $D_{j^*}W^*$.  Hence we have
\be
2K_{i^*j}\,\p_{\bar z}v^j = -e^{-\phi}e^{K/2}D_{i^*}W^*  \; .
\ee
Finally, contracting with the inverse metric $K^{ki^*}$ and relabelling indices gives the result,
\be
2i\,\p_{\bar z}v^i = -ie^{-\phi}e^{K/2}K^{ij^*}D_{j^*}W^* \; ,
\ee
which coincides with \eqref{eq:T-goal} when $\p_{\bar z}b^i = 0$. 

\subsubsection*{Axion constraint}

To conclude the matching, let us derive the axion constraint \eqref{eq:KSEaxion}.  Here we will show that it is nothing other than the additional constraint \eqref{eq:extra10dconstraint} we imposed to complete the dilaton matching, namely,
\begin{equation}
\int\p_z\Omega\wedge\bar\Omega = \int\Omega\wedge\p_z\bar\Omega \; . \label{eq:KSEaxion10d}
\end{equation}

To prove that this matches \eqref{eq:KSEaxion}, we should expand it in terms of basis forms.  We will make use of the prepotential in the large complex strucutre limit,
\begin{equation}
\cG = -\frac{1}{6}\tcK_{abc}\frac{\cZ^a\cZ^b\cZ^c}{\cZ^0} \; ,
\end{equation}
where $\cZ^A = \cZ^0(1,Z^a)$ and $\cG_A = \p\cG /\p\cZ^A$.  Explicitly, we find that
\begin{equation}
\cG_a = -\frac{\cZ^0}{2}\tcK_{abc}Z^bZ^c \; , \;\;\;
\cG_0 = \frac{\cZ^0}{6}\tcK_{abc}Z^aZ^bZ^c \; . \label{eq:G-lcs}
\end{equation}

Now use \eqref{expansion} to rewrite \eqref{eq:KSEaxion10d} as
\begin{equation}
\barcZ^A\left(\p_z\cG_A\right) - \left(\p_z\cZ^A\right)\barcG_A = \left(\p_z\barcZ^A\right)\cG_A - \cZ^A\left(\p_z\barcG_A\right) \; . \label{eq:KSEaxion10dexp}
\end{equation}
In the large complex structure limit, we can use \eqref{eq:G-lcs} to expand the left-hand side of this expression as
\bea
\barcZ^A\left(\p_z\cG_A\right) - \left(\p_z\cZ^A\right)\barcG_A 
&=& \frac{\left(\cZ^0\right)^2}{2}\tcK_{abc}\p_zZ^a\left[Z^bZ^c - 2Z^b\barZ^c + \barZ^b\barZ^c\right] \nn \\
&=& -2\left(\cZ^0\right)^2\tcK_a\p_zZ^a \; . \label{eq:axion10dleft}
\eea
To obtain the second equality we have used the \eqref{metric2} and the symmetry of $\tcK_{abc}$ under permutations of the indices.  Similarly, the right-hand side can be expanded to give
\be
\left(\p_z\barcZ^A\right)\cG_A - \cZ^A\left(\p_z\barcG_A\right) 
= 2\left(\cZ^0\right)^2\tcK_a\p_z\barZ^a \; . \label{eq:axion10dright}
\ee
Finally, equating \eqref{eq:axion10dleft} and \eqref{eq:axion10dright}, we find
\be
-2\left(\cZ^0\right)^2\tcK_a\p_zZ^a = 2\left(\cZ^0\right)^2\tcK_a\p_z\barZ^a \; ,
\ee
which can be rearranged to give
\be
\tcK_a\p_zc^a = 0 \; .
\ee
This expression matches \eqref{eq:KSEaxion} when $\p_z a = \p_z b^i = 0$.  Thus we have completed the matching.


\section{Discussion and outlook}
\label{sec:disc}
In this paper we have investigated compactifications of heterotic string theory on eight-dimensional geometries with two non-compact directions. We have performed the matching between the ten-dimensional geometries and the four-dimensional BPS equations that generically correspond to 1/4-BPS solutions. In doing so, we have established a connection between eight-dimensional $Spin(7)$ structures and 1/4-BPS vacua.   

In addition, we have introduced new types of 1/2-BPS solutions corresponding to topological cosmic strings and black holes (see appendix \ref{sec:csandbh}). We have not attempted to construct explicit solutions to the more general 1/4-BPS equations. It is expected that such solutions will generically break more spacetime symmetry than the usual 1/2-BPS solutions of e.g. spherical or cylindrical symmetry as given in appendix \ref{sec:csandbh}. One could for example look for cosmic string solutions where the radial symmetry is preserved but with broken axial symmetry, or vice versa. Solutions of the form of intersecting branes also come to mind \cite{Carroll:1999wr, Gauntlett:2000ch}. We will explore these directions in a future publication \cite{Angus2015}.

As we have seen, these types of compactifications allow for a more generic superpotential than what is obtainable in the previous domain wall compactifications. In particular, 
the flow equations allow the internal manifold to be a \emph{generalized half-flat manifold}. The resulting superpotential takes a form very similar to those of type II non-geometric compactifications \cite{shelton2005nongeometric, aldazabal2006more, Grana:2006hr, micu2007towards}, and it would be interesting to study this relation further. In particular, it would be interesting to look for compactifications which might be suitable for inflationary models, as has been done on the type II side \cite{hassler2014inflation, blumenhagen2015towards}.   

This paper has mostly been concerned with establishing the matching between the ten-dimensional and four-dimensional solutions of the BPS equations. However, one important task remains, namely, to study explicit examples of solutions, as has been done for the domain wall case in e.g. \cite{Klaput:2011mz, Gray:2012md, Klaput:2012vv, Klaput:2013nla}. Furthermore, in order to study the phenomenology of these solutions, they have to be lifted to maximally symmetric vacua by means of non-perturbative effects such as gaugino condensates \cite{Klaput:2012vv, Chatzistavrakidis:2012qb}, or one can look for non-supersymmetric perturbative vacua as was done in \cite{Lukas:2015kca}. Work in this direction is in progress.


\acknowledgments
We thank Andre Lukas and Michael Klaput for very early collaboration on this topic, as well as Vishnu Jejjala for enjoyable discussion on domain walls. 
This work was supported by IBS under the project code IBS-R018-D1. EES is supported by the ILP LABEX (under reference ANR-10-LABX-63), and by French
state funds managed by the ANR within the Investissements dAvenir program under reference
ANR-11-IDEX-0004-02.
CM is indebted to the National Institute for Theoretical Physics, the Mandelstam Institute for Theoretical Physics, and the University of the Witwatersrand for hospitality and financial support.

\appendix
\section*{Appendices}

\section{K\"ahler and complex structure moduli}\label{moduli}

In this appendix we will review some useful formulae that can be derived from the properties of Calabi-Yau moduli spaces. All of the following relations are assumed to hold also in the context of our ansatz for $SU(3)$-structure manifolds, the only difference being that the basis forms are no longer harmonic. 

Let us start with the two-form $J$ and its corresponding moduli field. The volume of the Calabi-Yau (or $SU(3)$ structure manifold) $X$ is given by
\be
  {\cal V}=\frac{1}{6}\int_XJ\wedge J\wedge J \;. \label{eq:VJJJ}
\ee
The metric appearing in the deformation space of the metric of $X$ is given by
\begin{equation}\label{Jkahlermetric}
K^{\left(T\right)}_ {ij}=\frac{1}{4\mathcal{V}}\int_X\omega_i\wedge*\omega_j\;,
\end{equation}
where
\be
  *\omega_i=-J\wedge\omega_i+\frac{3}{2}\,\frac{\int_X J\wedge J\wedge \omega_i}{\int_X J\wedge J\wedge J}\,J\wedge J \;.
\ee
The K\"{a}hler potential thus corresponds to
\be
  K^{\left(T\right)}_{ij}=\frac{\partial^2K^{\left(T\right)}}{\partial T^i\partial\bar{T}^j} \;, \;\;\; K^{\left(T\right)}=-\ln\left(\frac{4}{3}\int_XJ\wedge J\wedge J\right) \;,
\ee
which is the logarithm of the volume of $X$.

These expressions can be simplified by introducing more convenient notation. Let us write the triple intersection numbers,
\begin{equation}
  {\cal K}_{ijk}=\int_X\omega_i\wedge\omega_j\wedge\omega_k \; ,
\end{equation}
and define contractions with the moduli fields $v^i$,
\begin{equation}\label{metric1}
  {\cal K}={\cal K}_{ijk}v^iv^jv^k \;, \;\;\; {\cal K}_i={\cal K}_{ijk}v^jv^k \;, \;\;\; {\cal K}_{ij}={\cal K}_{ijk}v^k \;.
\end{equation}
The K\"ahler metric~\eqref{Jkahlermetric} can thus be re-written as
\begin{equation}
  K^{\left(T\right)}_ {ij}=\frac{9}{4}\frac{\mathcal{K}_i\mathcal{K}_j}{\mathcal{K}^2}-\frac{3}{2}\frac{\mathcal{K}_{ij}}{\mathcal{K}} \;, \quad {K^{\left(T\right)}}^{ij}=-\frac{2}{3}{\cal K}\left({\cal K}^{ij}-3\frac{v^iv^j}{{\cal K}}\right) \; , \label{eq:Kij_inverse}
\end{equation}
where we also gave the inverse metric ${K^{\left(T\right)}}^{ij}$. Here the coefficients ${\cal K}^{ij}$ are defined from the property ${\cal K}^{ij}{\cal K}_{jk}=\delta^i_{\phantom {i}k}$.
Another useful formula is the contraction with the first-order derivative $K^{(T)}_i$,
\be
  {K^{\left(T\right)}}^{ij}K^{(T)}_j=-2iv^i \; , \;\;\; {\rm where} \;\;\; K^{\left(T\right)}_i\equiv \frac{\partial K^{\left(T\right)}}{\partial T^i}=\frac{3i}{2}\frac{\mathcal{K}_i}{\mathcal{K}} \;. \label{eq:Ki_contract}
\ee

The properties of K\"ahler and complex structure moduli have striking similarities, a fact that led to the conjecture of mirror symmetry. For each Calabi-Yau $X$, there exists a mirror Calabi-Yau $\tilde X$ whose K\"ahler and complex structure moduli are exchanged. This conjecture allows us to introduce (triple) intersections numbers of the mirror Calabi-Yau $\tilde X$,
\begin{equation}
  \tilde{\mathcal{K}}=\tilde{\mathcal{K}}_{abc}w^aw^bw^c \;, \;\;\; \tilde{\mathcal{K}}_a=\tilde{\mathcal{K}}_{abc}w^bw^c \;, \;\;\; \tilde{\mathcal{K}}_{ab}=\tilde{\mathcal{K}}_{abc}w^c \;, \label{metric2}
\end{equation}
where we also wrote the relevant contractions. Mirror symmetry leads to the same expressions for the K\"ahler metric for the deformations of the complex structure as in the case of the  K\"ahler moduli space,
\begin{equation}
  K^{\left(Z\right)}_{a\bar b}=\frac{9}{4}\frac{\tilde{\mathcal{K}}_a\tilde{\mathcal{K}}_b}{\tilde{\mathcal{K}}^2}-\frac{3}{2}\frac{\tilde{\mathcal{K}}_{ab}}{\tilde{\mathcal{K}}} \;,\quad {K^{\left(Z\right)}}^{ab}=-\frac{2}{3}\tilde{{\cal K}}\left(\tilde{{\cal K}}^{ab}-3\frac{w^aw^b}{\tilde{{\cal K}}}\right) \;,
\end{equation}
where by definition $\tilde{{\cal K}}^{ab}\tilde{{\cal K}}_{bc}=\delta^a_{\phantom {a}c}$. The following relation for the first order derivative,
\be
  {K^{\left(2\right)}}^{ab}K_b=-2iw^a \quad {\rm where} \quad K^{\left(2\right)}_{a}=\frac{\partial K^{\left(2\right)}}{\partial Z^a}=\frac{3i}{2}\frac{\tilde{\mathcal{K}}_a}{\tilde{\mathcal{K}}} \;
\ee
can also easily be verified.

Finally, let us derive a useful result.  The volume of $X$ can be expressed in terms of the complex structure moduli as
\begin{equation}
\cV = \frac{i}{\vert\vert\Omega\vert\vert^2}\int_X\Omega\wedge\bar{\Omega} \; . \label{eq:VOO}
\end{equation}
Using \eqref{eq:VJJJ} and \eqref{eq:VOO}, the full K\"{a}hler potential defined in \eqref{hfk} and \eqref{K} can be written in the form
\begin{equation}
K = -\ln\left(16\cV^2\vert\vert\Omega\vert\vert^2e^{-2\phi}\right) \; ,
\end{equation}
from which it follows that
\begin{equation}
e^{-\phi}e^{K/2} = \frac{1}{4\vert\vert\Omega\vert\vert\cV} \; . \label{eq:K4OV}
\end{equation}

\section{$1/2$-BPS states}
\label{sec:csandbh}
In this paper we have focused on $1/4$-BPS solutions.  However, another possibility suggested by the general structure of \eqref{eq:KSEaA} and \eqref{eq:KSEbB} is that there should also exist $1/2$-BPS states other than the domain wall solution detailed in \cite{lukas2011g}.  Such solutions will admit a $G_2$ structure in which the six-dimensional $SU(3)$ structure is fibred over an interval that is not a Euclidean coordinate direction.  In this appendix we explore a couple of possibilities.

\subsection{Cosmic strings}

First let us consider an alternative solution to \eqref{eq:KSEaA} and \eqref{eq:KSEbB} whose holonomy corresponds to a fibration of $SU(3)$ over the radial direction in polar coordinates. To this end we impose that the warp factor, as well as scalar and spinor fields, should depend only on the perpendicular distance $\rho = \sqrt{x^2+y^2}$ from the origin, such that the solution is rotationally invariant around the $z$-axis. This ansatz will lead to solutions that are topological cosmic strings.

To obtain such solutions, note that we can satisfy (\ref{eq:KSEaA}) and (\ref{eq:KSEbB}) 
by choosing \hbox{$B = B(\rho)$,} $A^I = A^I(\rho)$, and $\zeta = \zeta(\rho)$.  This simplifies the Killing spinor equations to
\begin{align}
A^I_\rho &= -ie^{-B}e^{K/2}K^{IJ^*}D_{J*}W^* \; , \nn \\
B_\rho &= ie^{-B}e^{K/2}W \; , \nn \\
0 &= \text{Im}(K_IA^I_\rho) \; , \nn \\
2\zeta_\rho &= -B_\rho\zeta \; , \label{eq:KSE4d-rived}
\end{align}
where the $\rho$ subscript denotes a derivative with respect to $\rho$. Furthermore, this requires that the spinor $\zeta$ satisfies the constraint
\begin{equation}
\overline{\zeta} = \sigma^{\underline{\rho}}\zeta \; ,
\end{equation}
where we have defined
\begin{equation}
\sigma^{\underline{\rho}} = \frac{x}{\rho}\sigma^{\underline{1}} + \frac{y}{\rho}\sigma^{\underline{2}}
= \left(\begin{array}{c c}
0 & e^{-i\varphi} \\ e^{i\varphi} & 0
\end{array}\right)
\; ,
\end{equation}
for azimuthal angle $\varphi$.  Note that equations \eqref{eq:KSE4d-rived} are parametrically identical to the Killing spinor equations for the domain wall case described in \cite{lukas2011g}, so we expect the solutions to take the same form but with the $y$-direction substituted for the $\rho$-coordinate.

\subsection{Black holes}
We can generalize the above construction to three spatial dimensions.  The metric is now
\begin{equation}
ds_4^2=e^{-2B}\left(-dt^2 +\delta_{ij}\d x^i \d x^j\right) \label{eq:4dBHmetric} \; ,
\end{equation}
where we allow $B$ to depend on all three spatial directions $x^i$, with $\{i,j\} = \{1,2,3\}$.
For a naked singularity at the origin (a ``black hole'' solution) we would expect spherical symmetry, with the $SU(3)$ structure fibred over an interval corresponding to the radial distance $r$ from the singularity.  The spin connection is
\begin{equation}
\omega_0 = \frac{1}{2}B_i\sigma^{\underline{i}} \; , \;\;\;
\omega_i = i\frac{1}{2}\epsilon_{ij}^{\phantom{ij}k}B_k\sigma^{\underline{j}} \; .
\end{equation}

Maintaining a dependence on all three spatial coordinates for now, the Killing spinor equations \eqref{eq:KSE4d1} and \eqref{eq:KSE4d2} become
\begin{align}
A^I_i\sigma^{\underline{i}}\bar{\zeta} &= -ie^{-B}e^{K/2}K^{IJ^*}D_{J^*}W^*\zeta \; , \nn \\
B_i\zeta &= ie^{-B}e^{K/2}W\bar{\zeta} \; , \nn \\
\text{Im}(K_IA^I_i) &= 0 \; , \nn \\
2\zeta_i &= -B_i \; . \label{eq:BHconstraints}
\end{align}
Specializing to the scenario where the fields and warp factor are independent of the zenith and azimuthal angles $\theta$ and $\phi$, respectively, gives
\begin{align}
A^I_r &= -ie^{-B}e^{K/2}K^{IJ^*}D_{J*}W^* \; , \nn \\
B_r &= ie^{-B}e^{K/2}W \; , \nn \\
0 &= \text{Im}(K_IA^I_r) \; , \nn \\
2\zeta_r &= -B_r\zeta \; , \label{eq:KSE4d-BH}
\end{align}
where $r = \sqrt{x_ix^i}$ and the spinor satisfies
$\overline{\zeta} = \sigma^{\underline{r}}\zeta$, where
\begin{equation}
\sigma^{\underline{r}} = \frac{x_i}{r}\sigma^{\underline{i}}
= \left(\begin{array}{c c}
\cos\theta & \sin\theta\, e^{-i\phi} \\ \sin\theta\, e^{i\phi} & \cos\theta
\end{array}\right) \; .
\end{equation}




\begin{thebibliography}{10}

\bibitem{Candelas:1985en}
P.~Candelas, G.~T. Horowitz, A.~Strominger, and E.~Witten, {\it {Vacuum
  Configurations for Superstrings}},  {\em Nucl.Phys.} {\bf B258} (1985)
  46--74.

\bibitem{Berglund:1995yu}
P.~Berglund, P.~Candelas, X.~de~la Ossa, E.~Derrick, J.~Distler, et~al., {\it
  {On the instanton contributions to the masses and couplings of E(6)
  singlets}},  {\em Nucl.Phys.} {\bf B454} (1995) 127--163,
  [\href{http://arxiv.org/abs/hep-th/9505164}{{\tt hep-th/9505164}}].

\bibitem{Donagi:2004ub}
R.~Donagi, Y.-H. He, B.~A. Ovrut, and R.~Reinbacher, {\it {The Spectra of
  heterotic standard model vacua}},  {\em JHEP} {\bf 06} (2005) 070,
  [\href{http://arxiv.org/abs/hep-th/0411156}{{\tt hep-th/0411156}}].

\bibitem{Braun:2005bw}
V.~Braun, Y.-H. He, B.~A. Ovrut, and T.~Pantev, {\it {A Standard model from the
  E(8) x E(8) heterotic superstring}},  {\em JHEP} {\bf 06} (2005) 039,
  [\href{http://arxiv.org/abs/hep-th/0502155}{{\tt hep-th/0502155}}].

\bibitem{Braun:2005nv}
V.~Braun, Y.-H. He, B.~A. Ovrut, and T.~Pantev, {\it {The Exact MSSM spectrum
  from string theory}},  {\em JHEP} {\bf 05} (2006) 043,
  [\href{http://arxiv.org/abs/hep-th/0512177}{{\tt hep-th/0512177}}].

\bibitem{Braun:2006me}
V.~Braun, Y.-H. He, and B.~A. Ovrut, {\it {Yukawa couplings in heterotic
  standard models}},  {\em JHEP} {\bf 04} (2006) 019,
  [\href{http://arxiv.org/abs/hep-th/0601204}{{\tt hep-th/0601204}}].

\bibitem{Ambroso:2009jd}
M.~Ambroso and B.~Ovrut, {\it {The B-L/Electroweak Hierarchy in Heterotic
  String and M-Theory}},  {\em JHEP} {\bf 10} (2009) 011,
  [\href{http://arxiv.org/abs/0904.4509}{{\tt arXiv:0904.4509}}].

\bibitem{Anderson:2011ns}
L.~B. Anderson, J.~Gray, A.~Lukas, and E.~Palti, {\it {Two Hundred Heterotic
  Standard Models on Smooth Calabi-Yau Threefolds}},  {\em Phys. Rev.} {\bf
  D84} (2011) 106005, [\href{http://arxiv.org/abs/1106.4804}{{\tt
  arXiv:1106.4804}}].

\bibitem{Anderson:2012yf}
L.~B. Anderson, J.~Gray, A.~Lukas, and E.~Palti, {\it {Heterotic Line Bundle
  Standard Models}},  {\em JHEP} {\bf 06} (2012) 113,
  [\href{http://arxiv.org/abs/1202.1757}{{\tt arXiv:1202.1757}}].

\bibitem{Ovrut:2012wg}
B.~A. Ovrut, A.~Purves, and S.~Spinner, {\it {Wilson Lines and a Canonical
  Basis of SU(4) Heterotic Standard Models}},  {\em JHEP} {\bf 11} (2012) 026,
  [\href{http://arxiv.org/abs/1203.1325}{{\tt arXiv:1203.1325}}].

\bibitem{Anderson:2010mh}
L.~B. Anderson, J.~Gray, A.~Lukas, and B.~Ovrut, {\it {Stabilizing the Complex
  Structure in Heterotic Calabi-Yau Vacua}},  {\em JHEP} {\bf 1102} (2011) 088,
  [\href{http://arxiv.org/abs/1010.0255}{{\tt arXiv:1010.0255}}].

\bibitem{Anderson:2011ty}
L.~B. Anderson, J.~Gray, A.~Lukas, and B.~Ovrut, {\it {The Atiyah Class and
  Complex Structure Stabilization in Heterotic Calabi-Yau Compactifications}},
  {\em JHEP} {\bf 1110} (2011) 032, [\href{http://arxiv.org/abs/1107.5076}{{\tt
  arXiv:1107.5076}}].

\bibitem{Anderson:2014xha}
L.~B. Anderson, J.~Gray, and E.~Sharpe, {\it {Algebroids, Heterotic Moduli
  Spaces and the Strominger System}},  {\em JHEP} {\bf 1407} (2014) 037,
  [\href{http://arxiv.org/abs/1402.1532}{{\tt arXiv:1402.1532}}].

\bibitem{delaOssa:2014cia}
X.~de~la Ossa and E.~E. Svanes, {\it {Holomorphic Bundles and the Moduli Space
  of N=1 Supersymmetric Heterotic Compactifications}},  {\em JHEP} {\bf 1410}
  (2014) 123, [\href{http://arxiv.org/abs/1402.1725}{{\tt arXiv:1402.1725}}].

\bibitem{GarciaFernandez:2015hja}
M.~Garcia-Fernandez, R.~Rubio, and C.~Tipler, {\it {Infinitesimal moduli for
  the Strominger system and generalized Killing spinors}},
  \href{http://arxiv.org/abs/1503.07562}{{\tt arXiv:1503.07562}}.

\bibitem{delaOssa:2015maa}
X.~de~la Ossa, E.~Hardy, and E.~E. Svanes, {\it {The Heterotic Superpotential
  and Moduli}},  \href{http://arxiv.org/abs/1509.08724}{{\tt
  arXiv:1509.08724}}.

\bibitem{Anderson:2011cza}
L.~B. Anderson, J.~Gray, A.~Lukas, and B.~Ovrut, {\it {Stabilizing All
  Geometric Moduli in Heterotic Calabi-Yau Vacua}},  {\em Phys.Rev.} {\bf D83}
  (2011) 106011, [\href{http://arxiv.org/abs/1102.0011}{{\tt
  arXiv:1102.0011}}].

\bibitem{Maldacena:2000mw}
J.~M. Maldacena and C.~Nunez, {\it {Supergravity description of field theories
  on curved manifolds and a no go theorem}},  {\em Int. J. Mod. Phys.} {\bf
  A16} (2001) 822--855, [\href{http://arxiv.org/abs/hep-th/0007018}{{\tt
  hep-th/0007018}}].

\bibitem{Gauntlett:2002sc}
J.~P. Gauntlett, D.~Martelli, S.~Pakis, and D.~Waldram, {\it {G structures and
  wrapped NS5-branes}},  {\em Commun. Math. Phys.} {\bf 247} (2004) 421--445,
  [\href{http://arxiv.org/abs/hep-th/0205050}{{\tt hep-th/0205050}}].

\bibitem{Shahbazi:2015sba}
C.~S. Shahbazi, {\it {A class of non-geometric M-theory compactification
  backgrounds}},  \href{http://arxiv.org/abs/1508.01750}{{\tt
  arXiv:1508.01750}}.

\bibitem{lukas2011g}
A.~Lukas and C.~Matti, {\it {G-structures and Domain Walls in Heterotic
  Theories}},  {\em JHEP} {\bf 01} (2011) 151,
  [\href{http://arxiv.org/abs/1005.5302}{{\tt arXiv:1005.5302}}].

\bibitem{Klaput:2011mz}
M.~Klaput, A.~Lukas, and C.~Matti, {\it {Bundles over Nearly-Kahler Homogeneous
  Spaces in Heterotic String Theory}},  {\em JHEP} {\bf 09} (2011) 100,
  [\href{http://arxiv.org/abs/1107.3573}{{\tt arXiv:1107.3573}}].

\bibitem{Matti:2012roa}
C.~Matti, {\it {Generalized Compactification in Heterotic String Theory}},
  \href{http://arxiv.org/abs/1204.3247}{{\tt arXiv:1204.3247}}.

\bibitem{Gray:2012md}
J.~Gray, M.~Larfors, and D.~Lust, {\it {Heterotic domain wall solutions and
  SU(3) structure manifolds}},  {\em JHEP} {\bf 08} (2012) 099,
  [\href{http://arxiv.org/abs/1205.6208}{{\tt arXiv:1205.6208}}].

\bibitem{Klaput:2013nla}
M.~Klaput, A.~Lukas, and E.~E. Svanes, {\it {Heterotic Calabi-Yau
  Compactifications with Flux}},  {\em JHEP} {\bf 09} (2013) 034,
  [\href{http://arxiv.org/abs/1305.0594}{{\tt arXiv:1305.0594}}].

\bibitem{de2014exploring}
X.~de~la Ossa, M.~Larfors, and E.~E. Svanes, {\it {Exploring SU(3) Structure
  Moduli Spaces with Integrable G2 Structures}},
  \href{http://arxiv.org/abs/1409.7539}{{\tt arXiv:1409.7539}}.

\bibitem{Klaput:2012vv}
M.~Klaput, A.~Lukas, C.~Matti, and E.~E. Svanes, {\it {Moduli Stabilising in
  Heterotic Nearly K\"ahler Compactifications}},
  \href{http://arxiv.org/abs/1210.5933}{{\tt arXiv:1210.5933}}.

\bibitem{Chatzistavrakidis:2008ii}
A.~Chatzistavrakidis, P.~Manousselis, and G.~Zoupanos, {\it {Reducing the
  Heterotic Supergravity on nearly-Kahler coset spaces}},  {\em Fortsch. Phys.}
  {\bf 57} (2009) 527--534, [\href{http://arxiv.org/abs/0811.2182}{{\tt
  arXiv:0811.2182}}].

\bibitem{Chatzistavrakidis:2009mh}
A.~Chatzistavrakidis and G.~Zoupanos, {\it {Dimensional Reduction of the
  Heterotic String over nearly-Kaehler manifolds}},  {\em JHEP} {\bf 09} (2009)
  077, [\href{http://arxiv.org/abs/0905.2398}{{\tt arXiv:0905.2398}}].

\bibitem{Lechtenfeld:2010dr}
O.~Lechtenfeld, C.~Nolle, and A.~D. Popov, {\it {Heterotic compactifications on
  nearly Kahler manifolds}},  {\em JHEP} {\bf 09} (2010) 074,
  [\href{http://arxiv.org/abs/1007.0236}{{\tt arXiv:1007.0236}}].

\bibitem{Gemmer:2011cp}
K.-P. Gemmer, O.~Lechtenfeld, C.~Nolle, and A.~D. Popov, {\it {Yang-Mills
  instantons on cones and sine-cones over nearly Kahler manifolds}},  {\em
  JHEP} {\bf 09} (2011) 103, [\href{http://arxiv.org/abs/1108.3951}{{\tt
  arXiv:1108.3951}}].

\bibitem{Gemmer:2012pp}
K.-P. Gemmer, A.~S. Haupt, O.~Lechtenfeld, C.~Nšlle, and A.~D. Popov, {\it
  {Heterotic string plus five-brane systems with asymptotic $\mathrm{AdS}_3$}},
   {\em Adv. Theor. Math. Phys.} {\bf 17} (2013), no.~4 771--827,
  [\href{http://arxiv.org/abs/1202.5046}{{\tt arXiv:1202.5046}}].

\bibitem{Chatzistavrakidis:2012qb}
A.~Chatzistavrakidis, O.~Lechtenfeld, and A.~D. Popov, {\it {Nearly K\"ahler
  heterotic compactifications with fermion condensates}},  {\em JHEP} {\bf 04}
  (2012) 114, [\href{http://arxiv.org/abs/1202.1278}{{\tt arXiv:1202.1278}}].

\bibitem{Gemmer:2013ica}
K.-P. Gemmer and O.~Lechtenfeld, {\it {Heterotic $G_2$-manifold
  compactifications with fluxes and fermionic condensates}},  {\em JHEP} {\bf
  11} (2013) 182, [\href{http://arxiv.org/abs/1308.1955}{{\tt
  arXiv:1308.1955}}].

\bibitem{Haupt:2014ufa}
A.~S. Haupt, O.~Lechtenfeld, and E.~T. Musaev, {\it {Order $\alpha'$ heterotic
  domain walls with warped nearly K\"ahler geometry}},  {\em JHEP} {\bf 11}
  (2014) 152, [\href{http://arxiv.org/abs/1409.0548}{{\tt arXiv:1409.0548}}].

\bibitem{Angus2015}
S.~Angus, C.~Matti, and E.~E. Svanes, {\it In progress},
  \href{http://arxiv.org/abs/16xx.xxxxx}{{\tt arXiv:16xx.xxxxx}}.

\bibitem{Strominger1986253}
A.~Strominger, {\it Superstrings with torsion},  {\em Nuclear Physics B} {\bf
  274} (1986), no.~2 253 -- 284.

\bibitem{Hull:1986kz}
C.~Hull, {\it {Compactifications of the Heterotic Superstring}},  {\em
  Phys.Lett.} {\bf B178} (1986) 357.

\bibitem{slansky1981group}
R.~Slansky, {\it {Group Theory for Unified Model Building}},  {\em Phys. Rept.}
  {\bf 79} (1981) 1--128.

\bibitem{Gauntlett:2003cy}
J.~P. Gauntlett, D.~Martelli, and D.~Waldram, {\it {Superstrings with intrinsic
  torsion}},  {\em Phys.Rev.} {\bf D69} (2004) 086002,
  [\href{http://arxiv.org/abs/hep-th/0302158}{{\tt hep-th/0302158}}].

\bibitem{Agricola:2006tx}
I.~Agricola, {\it {The Srni lectures on non-integrable geometries with
  torsion}},  {\em Submitted to: Rend. Circ. Mat. Palermo} (2006)
  [\href{http://arxiv.org/abs/math/0606705}{{\tt math/0606705}}].

\bibitem{deCarlos:2005kh}
B.~de~Carlos, S.~Gurrieri, A.~Lukas, and A.~Micu, {\it {Moduli stabilisation in
  heterotic string compactifications}},  {\em JHEP} {\bf 0603} (2006) 005,
  [\href{http://arxiv.org/abs/hep-th/0507173}{{\tt hep-th/0507173}}].

\bibitem{Gurrieri:2004dt}
S.~Gurrieri, A.~Lukas, and A.~Micu, {\it {Heterotic on Half-Flat}},  {\em
  Phys.Rev.} {\bf D70} (2004) 126009,
  [\href{http://arxiv.org/abs/hep-th/0408121}{{\tt hep-th/0408121}}].

\bibitem{Gurrieri:2007jg}
S.~Gurrieri, A.~Lukas, and A.~Micu, {\it {Heterotic String Compactifications on
  Half-flat Manifolds. II.}},  {\em JHEP} {\bf 0712} (2007) 081,
  [\href{http://arxiv.org/abs/0709.1932}{{\tt arXiv:0709.1932}}].

\bibitem{d2005gauging}
R.~D'Auria, S.~Ferrara, M.~Trigiante, and S.~Vaula, {\it {Gauging the
  Heisenberg algebra of special quaternionic manifolds}},  {\em Phys. Lett.}
  {\bf B610} (2005) 147--151, [\href{http://arxiv.org/abs/hep-th/0410290}{{\tt
  hep-th/0410290}}].

\bibitem{tomasiello2005topological}
A.~Tomasiello, {\it {Topological mirror symmetry with fluxes}},  {\em JHEP}
  {\bf 06} (2005) 067, [\href{http://arxiv.org/abs/hep-th/0502148}{{\tt
  hep-th/0502148}}].

\bibitem{grana2006hitchin}
M.~Grana, J.~Louis, and D.~Waldram, {\it {Hitchin functionals in N=2
  supergravity}},  {\em JHEP} {\bf 01} (2006) 008,
  [\href{http://arxiv.org/abs/hep-th/0505264}{{\tt hep-th/0505264}}].

\bibitem{shelton2005nongeometric}
J.~Shelton, W.~Taylor, and B.~Wecht, {\it {Nongeometric flux
  compactifications}},  {\em JHEP} {\bf 10} (2005) 085,
  [\href{http://arxiv.org/abs/hep-th/0508133}{{\tt hep-th/0508133}}].

\bibitem{aldazabal2006more}
G.~Aldazabal, P.~G. Camara, A.~Font, and L.~E. Ibanez, {\it {More dual fluxes
  and moduli fixing}},  {\em JHEP} {\bf 05} (2006) 070,
  [\href{http://arxiv.org/abs/hep-th/0602089}{{\tt hep-th/0602089}}].

\bibitem{Grana:2006hr}
M.~Grana, J.~Louis, and D.~Waldram, {\it {SU(3) x SU(3) compactification and
  mirror duals of magnetic fluxes}},  {\em JHEP} {\bf 0704} (2007) 101,
  [\href{http://arxiv.org/abs/hep-th/0612237}{{\tt hep-th/0612237}}].

\bibitem{micu2007towards}
A.~Micu, E.~Palti, and G.~Tasinato, {\it {Towards Minkowski Vacua in Type II
  String Compactifications}},  {\em JHEP} {\bf 03} (2007) 104,
  [\href{http://arxiv.org/abs/hep-th/0701173}{{\tt hep-th/0701173}}].

\bibitem{hassler2014inflation}
F.~Hassler, D.~Lust, and S.~Massai, {\it {On Inflation and de Sitter in
  Non-Geometric String Backgrounds}},
  \href{http://arxiv.org/abs/1405.2325}{{\tt arXiv:1405.2325}}.

\bibitem{blumenhagen2015towards}
R.~Blumenhagen, A.~Font, M.~Fuchs, D.~Herschmann, and E.~Plauschinn, {\it
  {Towards Axionic Starobinsky-like Inflation in String Theory}},  {\em Phys.
  Lett.} {\bf B746} (2015) 217--222,
  [\href{http://arxiv.org/abs/1503.01607}{{\tt arXiv:1503.01607}}].

\bibitem{kodaira1958deformations}
K.~Kodaira and D.~C. Spencer, {\it On deformations of complex analytic
  structures, i},  {\em Annals of Mathematics} (1958) 328--401.

\bibitem{Carroll:1999wr}
S.~M. Carroll, S.~Hellerman, and M.~Trodden, {\it {Domain wall junctions are
  1/4 - BPS states}},  {\em Phys. Rev.} {\bf D61} (2000) 065001,
  [\href{http://arxiv.org/abs/hep-th/9905217}{{\tt hep-th/9905217}}].

\bibitem{Gauntlett:2000ch}
J.~P. Gauntlett, G.~W. Gibbons, C.~M. Hull, and P.~K. Townsend, {\it {BPS
  states of D = 4 N=1 supersymmetry}},  {\em Commun. Math. Phys.} {\bf 216}
  (2001) 431--459, [\href{http://arxiv.org/abs/hep-th/0001024}{{\tt
  hep-th/0001024}}].

\bibitem{Lukas:2015kca}
A.~Lukas, Z.~Lalak, and E.~E. Svanes, {\it {Heterotic Moduli Stabilisation and
  Non-Supersymmetric Vacua}},  {\em JHEP} {\bf 08} (2015) 020,
  [\href{http://arxiv.org/abs/1504.06978}{{\tt arXiv:1504.06978}}].

\end{thebibliography}

\providecommand{\href}[2]{#2}\begingroup\raggedright\endgroup


\end{document}